\documentclass[prl,10pt,twocolumn,superscriptaddress]{revtex4}
\pdfoutput=1
\usepackage{graphicx}
\usepackage{floatflt}
\graphicspath{{figures/}{figures2/}}
\begin{document}

%%%%%%%%%%%%%%%%%%%%%%%
% Define new commands %
%%%%%%%%%%%%%%%%%%%%%%%
\newcommand{\oscdemod}{oscillator/demodulator}
\newcommand{\fMUX}{fMUX}
\newcommand{\lisa}{LISA}
\newcommand{\squid}{SQUID}
\newcommand{\rtHz}{$\sqrt{\mbox{Hz}}$}
\newcommand{\phinot}{\mbox{$\Phi_0$}}
\newcommand{\degree}{ \mbox{$^{\circ}$} }
\newcommand{\fortran}{{\tt Fortran~77}}
\newcommand{\CXX}{C++}
\newcommand{\order}{\mbox{${\cal O}$}}
\newcommand{\const}{\mbox{\sc\small Const}}
\newcommand{\mycomment}[1]{\marginpar{\it\tiny #1}} % used for margin comments
%\newcommand{\mycomment}[1]{\typeout{#1}}
%--------------------------------------------------------------

\title{Linearized SQUID Array (LISA) for High Bandwidth Frequency-Domain
Readout Multiplexing}

\author{T. M. Lanting}
\email{tlanting@dwavesys.com}
\affiliation{Physics Department, McGill University, Montreal, Canada H2T 2Y8}
\affiliation{D-Wave Systems Inc., 100-4401 Still Creek Drive, Burnaby, Canada, V5C 6G9}
\author{M. Dobbs}
\affiliation{Physics Department, McGill University, Montreal, Canada H2T 2Y8}
\author{H. Spieler}
\affiliation{Lawrence Berkeley National Laboratory, Berkeley, CA 94720}
\author{A.T. Lee}
\affiliation{Physics Department, University of California, Berkeley, CA 94720}
\author{Y. Yamamoto}
\affiliation{Physics Department, McGill University, Montreal, Canada H2T 2Y8}

\begin{abstract}
We have designed and demonstrated a Superconducting Quantum
Interference Device (SQUID) array linearized with cryogenic
feedback. To achieve the necessary loop gain a 300 element series
array SQUID is constructed from three monolithic 100-element series arrays.  A feedback resistor
completes the loop from the SQUID output to the input coil. The
short feedback path of this Linearized SQUID Array (LISA) allows for
a substantially larger flux-locked loop bandwidth as compared to a
SQUID flux-locked loop that includes a room temperature amplifier.
The bandwidth, linearity, noise performance, and 3 \phinot $\ $dynamic
range of the LISA are sufficient for its use in our target
application: the multiplexed readout of transition-edge sensor
bolometers.
\end{abstract}

%Uncomment for PACS numbers title message
%\pacs{00.00, 20.00, 42.10}
% Keywords required only for MST, PB, PMB, PM, JOA, JOB?
%\vspace{2pc}
%\noindent{\it Keywords}: Article preparation, IOP journals
% Uncomment for Submitted to journal title message
%\submitto{\JPA}
% Comment out if separate title page not required
\maketitle

%--------------------------------------------------------------
\section{Introduction}
\label{SEC:introduction}

\paragraph{}
A new generation of fully lithographed Transition-Edge Sensor (TES)
bolometers has been developed for astronomical observations
in the far-IR to millimeter wavelength range. These devices allow
for large arrays of $10^3 - 10^4$ or more bolometers(\cite{scuba2book,tran2007,grainger2008,ruhl2004,schwan2003}), providing a substantial step forward in
sensitivity. A significant challenge for scaling arrays is readout multiplexing.  We have developed the
superconducting quantum-interference device (SQUID) based
frequency-domain readout multiplexer
(\fMUX)~\cite{yoon01,Spieler02,2004NIMPA.520..548L,dobbs2007}. The
\fMUX\ is currently deployed on two active experiments
\cite{ruhl2004,schwan2003} and will be used for
a number of experiments in the near future
\cite{tran2007,oxley2004}. A digital implementation of the
room temperature components of the \fMUX\ \cite{dobbs2007} system is being integrated with these experiments. The
digital electronics has sufficient capacity to
extend the multiplexed channel count. A
complementary SQUID-based readout multiplexing technique is time
domain multiplexing
(\cite{chervenak99,2000NIMPA.444..107C,halpern2007}).
\paragraph{}
With the \fMUX\ system, the number of detectors that can be
instrumented with a single SQUID-based cryogenic amplifier is proportional to
the readout bandwidth. For the existing \fMUX\ system, the readout bandwidth of $\sim 1$ MHz allows multiplexing of 8-16 channels.  The readout bandwidth is fundamentally limited by the SQUID feedback electronics which include a stage at room temperature. In this
paper we report on the design and performance of the Linearized SQUID
Array (LISA) in which the first stage of the feedback electronics is moved
to the cryogenic stage. This configuration allows for an increase in
readout bandwidth and thus a corresponding increase in the number of multiplexed pixels by more than an order of magnitude.  
\paragraph{}
The LISA provides the cryogenic amplification
necessary to increase the channel count of the \fMUX\ system and coupled with
the digital room temperature electronics, this device represents a new generation in \fMUX\
technology. Furthermore, because of its simple design that is easily implemented on a single chip, its low power dissipation, and its low input noise current, the LISA is an excellent general purpose cryogenic transimpedance amplifier.
\paragraph{}
In this paper we describe the design and testing of a prototype LISA, utilizing several (three, in this example) 100-SQUID arrays connected in series with a cryogenic feedback resistor
connecting the combined output of the SQUIDs to the input coil. The
dynamic range, linearity, and input noise of the LISA meet the fMUX
design specifications across a readout bandwidth of 10 MHz.

%--------------------------------------------------------------
\section{Primary Application}
\label{SEC:primaryapplication}

\paragraph{}
Our target application for the \lisa\ is the multiplexed readout of
transition edge sensor (TES) bolometer arrays for the
detection of mm-wavelength radiation. These cryogenic detectors
employ a $\sim$3~mm metal absorber that is weakly coupled to a
$\sim$250~mK thermal bath. Radiation incident on the absorber
induces a temperature rise. This temperature change is measured with
a TES thermistor coupled to the absorber. The TES thermistor is 
operated in the transition between its superconducting state and its normal state where a small change in temperature yields a large change in resistance. Large arrays of
these TES bolometers can be manufactured photolithographically
making possible the new generation of mm-wavelength observations listed above.

\begin{figure}[h]
\centerline{\includegraphics[width=0.5\textwidth]{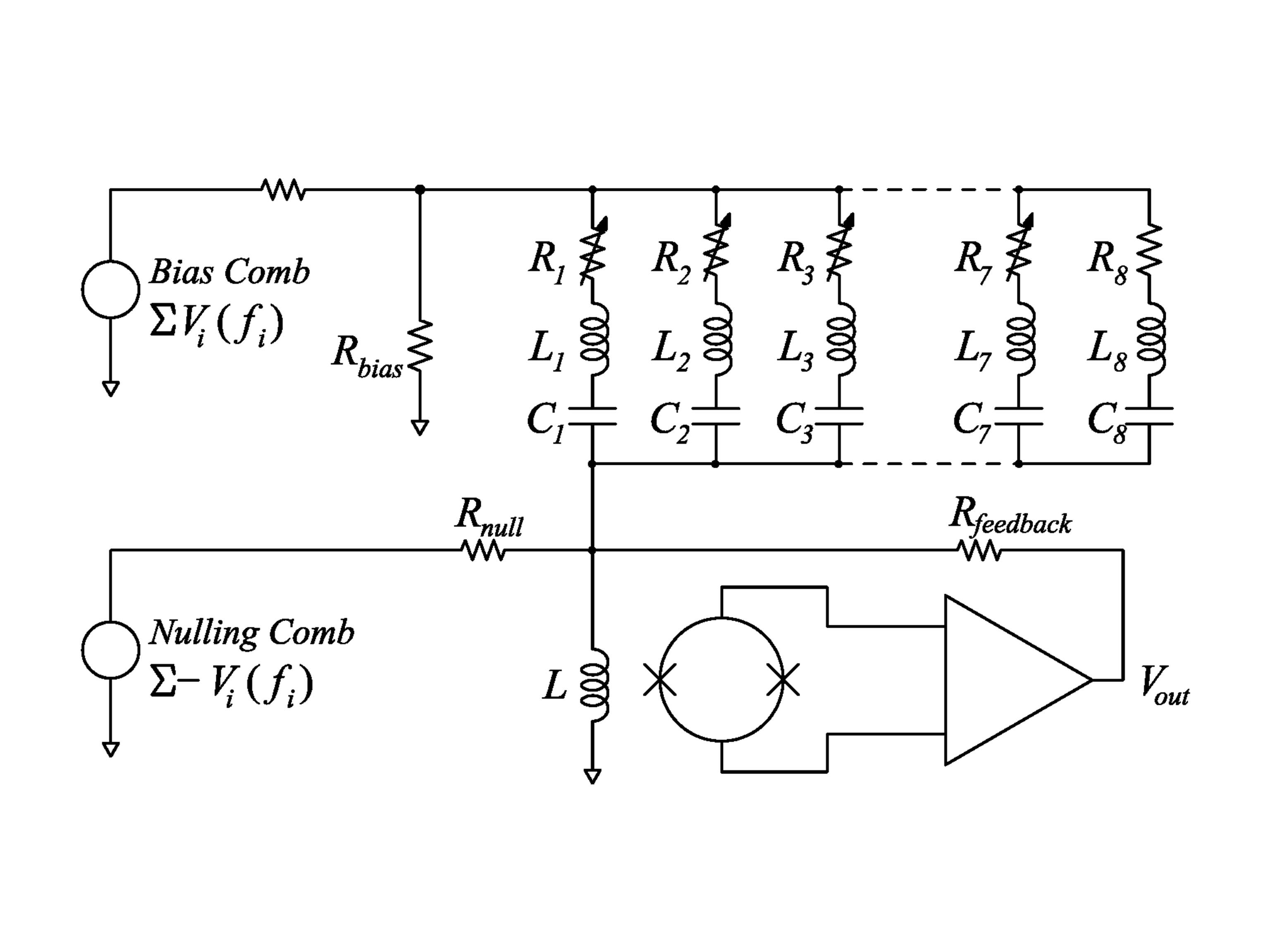}}
 \caption{\label{FIG:FDMschematic} Schematic of eight-channel \fMUX\  system.~\cite{2004NIMPA.520..548L}}
\end{figure}

\paragraph{}
In the \fMUX\ readout system, the TES devices are biased into their superconducting
transitions with high-frequency (0.3 -- 1 MHz) sinusoidal bias
voltages. A change in the incident radiation power induces a change
in the TES resistance that amplitude modulates its bias carrier. This translates the signal to sidebands above and below the 
carrier frequency. Each TES bolometer is biased at a different frequency, so the signals are now uniquely positioned in frequency space and the currents from the individual thermistors can be summed into one wire. The summed currents are fed to the input of a transimpedance amplifier utilizing a DC \squid\ operating in a flux-locked loop (FLL) configuration (see Fig. \ref{FIG:FDMschematic}). The TES thermistor typically has an impedance of $\sim 0.5 - 2\ \Omega$, a
noise current of 10-50~pA/\rtHz, and requires a bias voltage of
$\sim 5\ \mu\mbox{V}_{\rm{RMS}}$.

\paragraph{}
The SQUID electronics performance requirements for this application
are: (1) white noise lower than the TES noise ($< 10$ pA/\rtHz),
(2) sufficient bandwidth($\gg1$~MHz) to accommodate many carriers,
(3) input impedance much lower than the TES impedance ($\ll
0.5\Omega$) across the entire bandwidth to maintain constant-voltage bias, (4) sufficient
transimpedance ($Z_{\rm{LISA}}>150\Omega$) to override the
1~nV/\rtHz\ noise of a warm amplifier which follows the system, (5)
large ($> 10\ \mu\mbox{A}_{\rm{RMS}}$) dynamic range in order to
accommodate a single carrier signal, and (6) sufficient linearity to amplify large carrier signals and to reduce intermodulation between multiple carriers. Low
frequency noise in the SQUID electronics is not an issue as the
signals of interest have been translated to frequencies above 0.3~MHz.

\paragraph{}
The dynamic range requirement allows the system to handle at least
one full sized carrier.  This allows a user to tune the system in a
straight forward manner (tuning typically involves choosing the
appropriate bias voltage level for a particular sensor). After
tuning a single sensor, the carrier is nulled with a 180$^{\circ}$
unmodulated copy of the carrier signal injected at the input coil of
the SQUID. This does not affect the original signals, as all of their information is in the sidebands. Once the first carrier is nulled, a second one can be
added without increasing the dynamic range requirement. Thus, the
number of channels that can be multiplexed with a single set of
SQUID electronics is fundamentally limited by the bandwidth of these
electronics and not by the dynamic range.

%--------------------------------------------------------------
\section{Linearized SQUID Array}
\label{SEC:LISA}
\paragraph{}
The response of a SQUID has traditionally been linearized with a
feedback loop that includes a transistor amplifier operating at
300K. The bandwidth of this flux-locked loop circuit is limited by
propagation delays along the wires connecting this amplifier to the
SQUID which operates at the cold stage temperature ($\sim 4$K). The
current \fMUX\ bandwidth of 1 MHz is achieved by constraining the
wire length between 300K and 4K to be $<$ 0.15m. Further reductions
in wire length are not cryogenically feasible.

\paragraph{}
A significant increase in bandwidth can be achieved by either moving
the transistor amplifier to the cold stage or eliminating it
entirely from the feedback loop. The latter is preferable, since a
transistor amplifier would dissipate significant power on the cold
stage. With the amplifier removed from the circuit, a configuration
of SQUID devices can be used to produce the necessary loop gain
while maintaining the circuit's transimpedance.

\paragraph{}
One such configuration, referred to as the `SQUID op-amp', has been
demonstrated~\cite{SquidOpAmp}. That device uses a parallel cascade
of SQUIDs to increase the loop gain of the circuit. While each of
the SQUIDs in the parallel cascade contributes to the loop gain and
linearizes the circuit, the dynamic range of the circuit is limited
by the final stage SQUIDs. For the multiplexed readout of TES
bolometers, we require a configuration that extends both the linearity and
dynamic range of the SQUID system.

\begin{figure}[h]
\centerline{\includegraphics[width=0.6\textwidth]{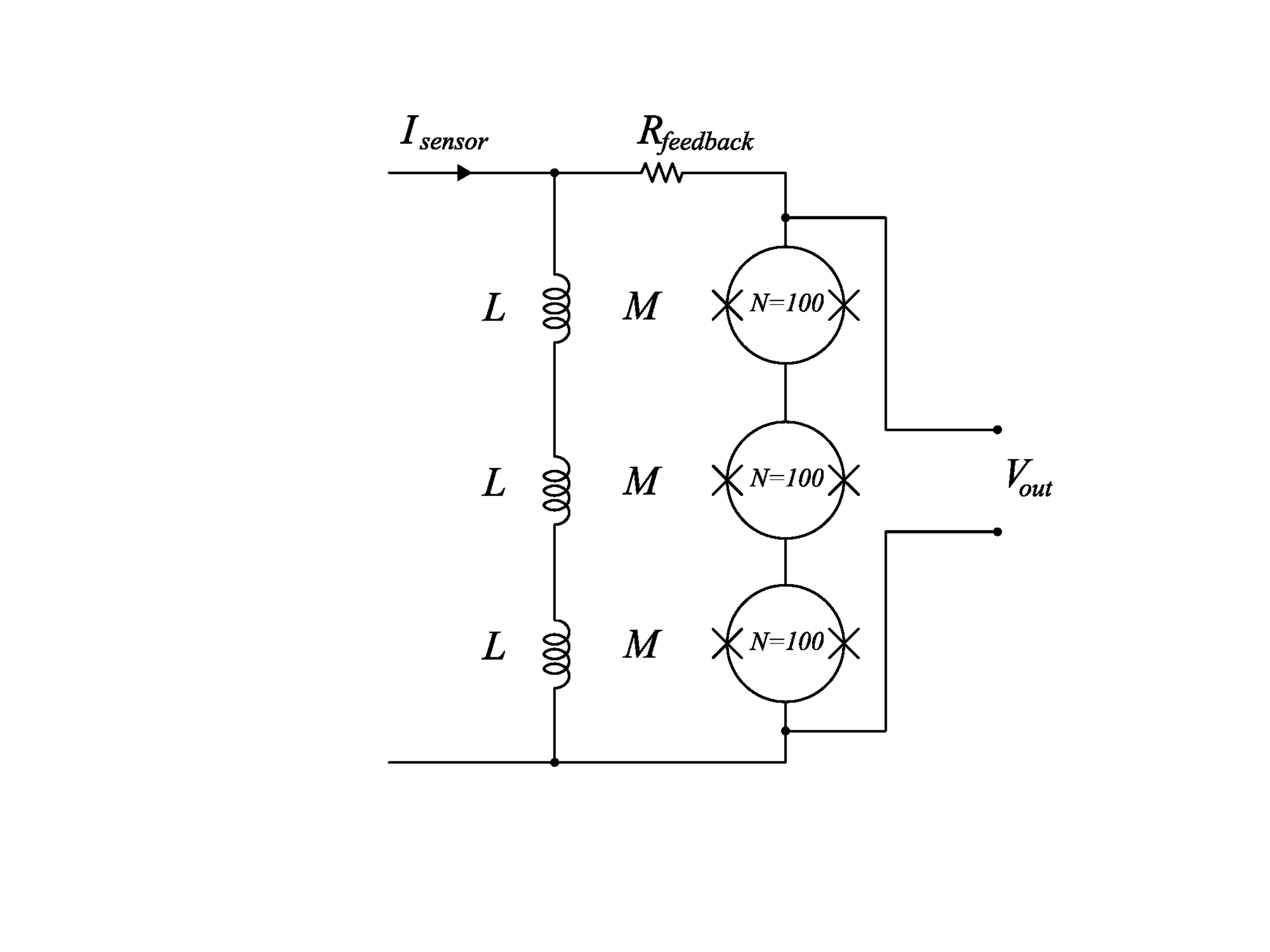}}
 \caption{\label{FIG:LISAschematic} Schematic of Linearized SQUID
   Array (LISA). Three one hundred element series array SQUIDs are
   connected in series. A feedback resistor connects the SQUID output to
   the input coil, completing the feedback loop.}
\end{figure}

\paragraph{}
To meet our requirements, we have developed and tested the
Linearized SQUID Array (LISA) concept
(Figure~\ref{FIG:LISAschematic}). The LISA eliminates the warm
transistor amplifier from the flux locked loop by using a series
configuration of SQUIDs to simultaneously linearize the circuit and
extend the dynamic range. The LISA prototype we discuss in this
paper consists of three monolithic series array SQUID chips \cite{NIST-arrays}, each consisting of
100~series-connected SQUID, themselves connected in series to form a 300
element series array. The output voltage of the LISA is coupled back
to the input coil with a feedback resistor to complete the feedback
loop (see figure \ref{FIG:LISAschematic}) and reduce the input impedance to meet the requirements of constant voltage bias to the thermistor.

\paragraph{}
In designing and testing any \squid\ FLL, we focus on four figures
of merit: the closed loop gain $A_{\rm{LISA}}$, the forward gain,
or transimpedance, $Z_{\rm{LISA}}$, the input noise current $i_n$,
and the closed loop bandwidth of the system. The closed loop gain
quantifies the ability of the feedback electronics to extend the
dynamic range and linearity of the SQUIDs that form the LISA. The
transimpedance measures the ratio of the LISA's output voltage to the input current. The input noise
current determines the sensitivity with which currents can be read
out with the LISA. The closed loop bandwidth determines the number
of sensors that can be multiplexed.

\paragraph{}
If the system consists of only an input-coil coupled \squid-like
component and a feedback resistor $R_{FB}$, the loop gain will be:
\begin{equation}
A_{\rm{LISA}}=\frac{Z_{0}}{R_{FB}+R_{0}} \label{EQN:loopgain}
\end{equation}
where $Z_{0}$ and $R_{0}$ are the transimpedance and output
impedance, respectively, of the series array SQUID combination shown
in figure \ref{FIG:LISAschematic}. The transimpedance is the product
of the mutual inductance between the input coil and the \squid\ inductance and the slope of the voltage response to applied flux:

\begin{equation}
Z_{0} = M\ \frac{\partial{V}}{\partial{\phi}}\mid_{\ \rm{max}}
\label{EQN:ZSQ}
\end{equation}

A current supplied to the input coil of the LISA produces an output
voltage $v_{LISA}=-i_{coil} Z_{LISA}$. The LISA transimpedance is then

\begin{eqnarray*}
Z_{LISA} &=& (R_{FB}+R_{0})  \\
& || &
Z_{0} \left( 1 - \frac{R_{0}}{R_{FB}+R_{0}} \right).
\label{EQN:transimpedance}
\end{eqnarray*}

The expected current noise is a quadrature sum of the intrinsic
noise from the component SQUID arrays, $i_{0}$, and the Johnson
noise of the feedback resistor, $i_{\rm{FB}}$:

\begin{equation}
i_n = \sqrt{i_{0}^2 + i_{\rm{FB}}^2} \label{EQN:currentnoise}
\end{equation}

\paragraph{}
Equations \ref{EQN:loopgain} and \ref{EQN:transimpedance} show a
fundamental limitation on performance that depends on the properties
of the individual SQUIDs that form the LISA. The output equivalent circuit is a voltage source with a series impedance $R_{0}$ inherent to the SQUIDs. The feedback current is then determined by the series combination of $R_{FB}$ and $R_{0}$ while the output voltage of the LISA depends on the voltage divider formed by the output resistance $R_{0}$ and the parallel combination of the feedback resistance $R_{FB}$ and the load presented by the capacitance of the cable connecting to the subsequent electronics at room temperature (assuming the warm amplifier's input impedance is sufficiently to be negligible). Thus,
for large output resistances $R_{0} \gg R_{FB}$ the loop gain
approaches the limit $Z_{0}/R_{0}$ and the transimpedance approaches
zero. In choosing or designing SQUIDs for the LISA, the ratio
$R_{0}/Z_{0}$ should be chosen to be as small as possible while
maintaining the required overall transimpedance, $Z_{\rm{LISA}}$. For a given load capacitance, $R_{0}$ also limits the achievable bandwidth.

%--------------------------------------------------------------
\section{Performance}
\label{SEC:performance}

\paragraph{} As previously mentioned, our prototype LISA was designed with three SQUID array
chips wired in series. The SQUID array on each chip has an input inductance of 160 nH, an input current noise of
2.5 pA/\rtHz, a mutual inductance $M = 80$ pH, a maximum
transimpedance $Z_{0}$ of $~ 450 \pm 50\ \Omega$, and an output impedance
$R_{0}$ of $80\pm 5\ \Omega$. We chose a feedback resistance of $R_{FB}
= 230\ \Omega$.

\subsection{Tuning}
Tuning a traditional SQUID device that includes room temperature
feedback electronics typically involves first choosing a SQUID
Josephson junction current bias that maximized the amplitude of the
voltage-flux relation, then choosing an appropriate flux bias to
maximize the \squid\ transimpedance. Finally, a switch is closed,
connecting the room temperature electronics output through a
feedback resistor to the input coil of the SQUID device, enabling
the flux-locked loop. The operating point of a flux-locked loop will jump by a single flux quantum if an input signal is applied that exceeds its dynamic range. Traditional flux-locked loops are restored to their original operating point by resetting the feedback either by interrupting the feedback signal momentarily and then re-locking or by turning the SQUID bias current to zero and then restoring it. 

\paragraph{}
Because of its design, the cryogenic connection between the LISA output and
input is inconvenient to open and close during the tuning
process. However, tuning is very straightforward for the LISA. First we varied the SQUID Josephson junction current bias and for each bias current setting we measured the DC voltage-flux relation of the LISA. We chose the current bias that maximizes the dynamic range of the LISA (this ensures that we are biasing the constituent SQUIDs at a point close to the maximum of their individual voltage-flux responses). Then we used the measured optimum voltage-flux relation to determine the flux bias signal that moves the LISA to the middle of its dynamic range and applied this flux to the device. If the operating point of the LISA jumped in response to a signal that exceeded its dynamic range, the LISA was reset by momentarily interrupting its current bias.

\subsection{Dynamic Range and Linearity}

\begin{figure}[h]
 \centerline{\includegraphics[width=0.5\textwidth]{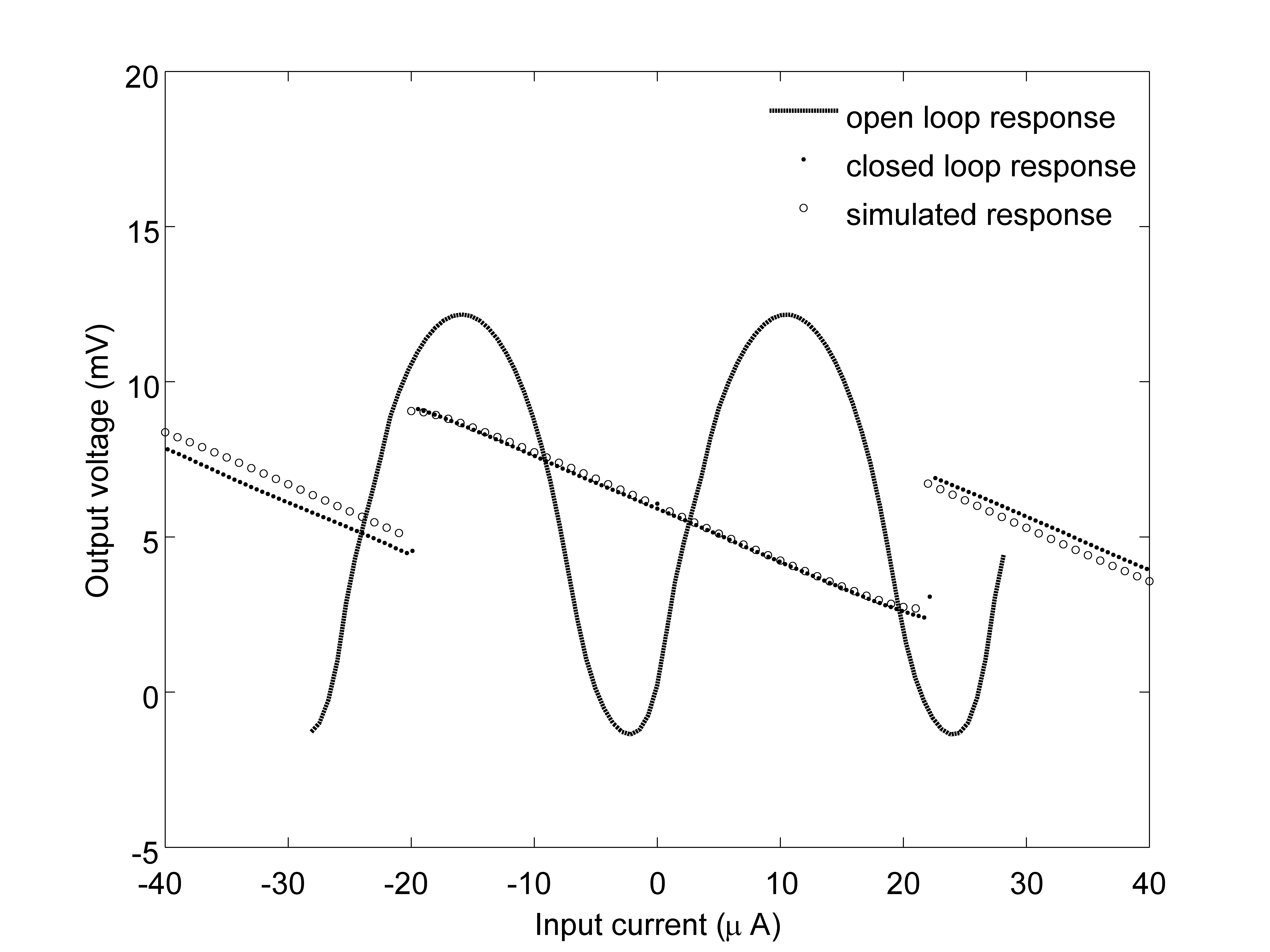}}
 \caption{\label{FIG:dynamicrange} LISA response to applied input current with and without application of cold feedback.}
\end{figure}

\paragraph{}
Once the LISA is tuned, we measure its dynamic
range (maximum peak-to-peak applied signal) by applying a series of DC currents to the input coil and measuring the output voltages
(voltage-current or voltage-flux relation). Figure
\ref{FIG:dynamicrange} shows this measured relation for three SQUID
devices in series without feedback, and the same relation for the
LISA. The simulated response is also plotted in figure \ref{FIG:dynamicrange}~\cite{lantingdissertation}. The intrinsic sinusoidal response has been linearized over a
dynamic range of $37 \pm 1 \ \mu A_{pp}$ and agrees well with the simulated response. Note that this measurement of dynamic range with a direct input current does not exactly apply to the high frequency response because of the 160 nH inductance of the input coil of each SQUID array. At 10 MHz, this corresponds to a reactance of $10\ \Omega$ from each SQUID array which slightly lowers the closed loop gain.

\paragraph{}
We expect that the dynamic range of the LISA, measured as a peak-to-peak current applied to the input coil, will be~\cite{Spieler02}

\begin{equation}
i_{pp} = \frac{\Phi_o}{M} \left(\frac{1}{2} + \frac{2 A_{\rm{LISA}}}{2\pi}\right),
\label{EQN:dynamicrange}
\end{equation}
where $\Phi_0$ is the quantum of magnetic flux, $M$ is the mutual inductance between the input coil and each one of the constituent SQUIDs, and $A_{loop}$ is the closed loop output impedance of the cold FLL
as defined in section \ref{SEC:LISA}. For the demonstration, we wired
three 100-element series array SQUID chips in series. For an arbitrary
number, $n$, of series array SQUID chips, the loop gain becomes:

\begin{equation}
A_{\rm{LISA}} = \frac{n Z_{\rm{single}}}{R_{FB} + n R_{\rm{single}}}
\end{equation}
where $Z_{\rm{single}}$ and $R_{\rm{single}}$ are the transimpedance
and output resistance of the individual series array SQUID devices.
As mentioned in section \ref{SEC:LISA}, for a large number of
devices in series, the loop gain approaches the limiting value of
$Z_{\rm{single}}/R_{\rm{single}} = 6.2$.

\paragraph{}
Using the properties of the individual LISA components used in this
demonstration, we expect $A_{loop} = 3.2$. This loop gain predicts,
using equation \ref{EQN:dynamicrange}, a dynamic range of 39 $\mu
A_{pp}$ in good agreement with measurement.

\begin{figure}[h]
 \centerline{\includegraphics[width=0.5\textwidth]{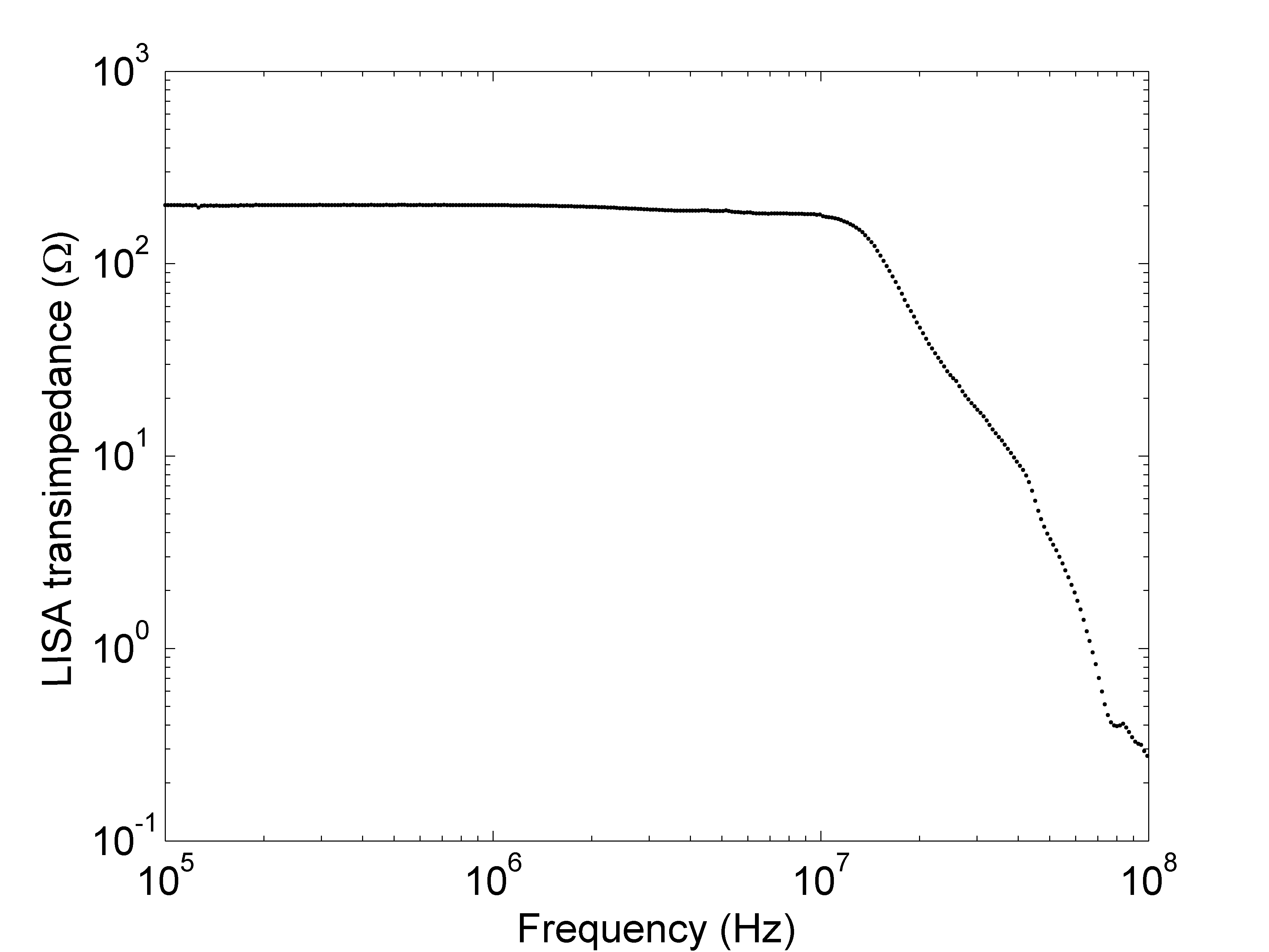}}

 \caption{\label{FIG:LISAbandwidth} Measured LISA transimpedance (using sinusoidal input of 2 $\mu$A$_{\rm{RMS}}$) over three decades of frequency.}
\end{figure}

The LISA has high linearity across most of its dynamic range, but exhibits gain suppression for large signals. We measure compression of $0.3\pm0.1$ dB at $\pm 19\ \mu$A with much smaller deviations from linearity for smaller signals.

\subsection{Bandwidth}
We measured the bandwidth of the LISA with a network analyser
~\cite{hp4195a}. We applied a current of $2\ \mu$A$_{\rm{RMS}}$ at the LISA input and
measured the output voltage across three decades of frequency (0.1 MHz to
100 MHz). The current was applied differentially to the input coil of the LISA and the output voltage was amplified with room temperature
electronics and then measured with a network analyser. We measure a small-signal
bandwidth of $14 \pm 1$ MHz (see figure \ref{FIG:LISAbandwidth}).  We measure the large-signal bandwidth by recording the size of input current signal needed to induce flux jumping in the LISA as a function of frequency. The large-signal bandwidth achieved with the prototype setup is $10$ MHz. 
For our setup, the measured bandwidth is currently limited by several room temperature components: a stray capacitance of 50 pF at the input of the room temperature amplifier and the room temperature amplifier itself. Using a well-matched room temperature amplifier with low input capacitance and a larger designed bandwidth will increase the available LISA bandwidth.

\subsection{Noise}
\begin{figure}[h]
 \centerline{\includegraphics[width=0.5\textwidth]{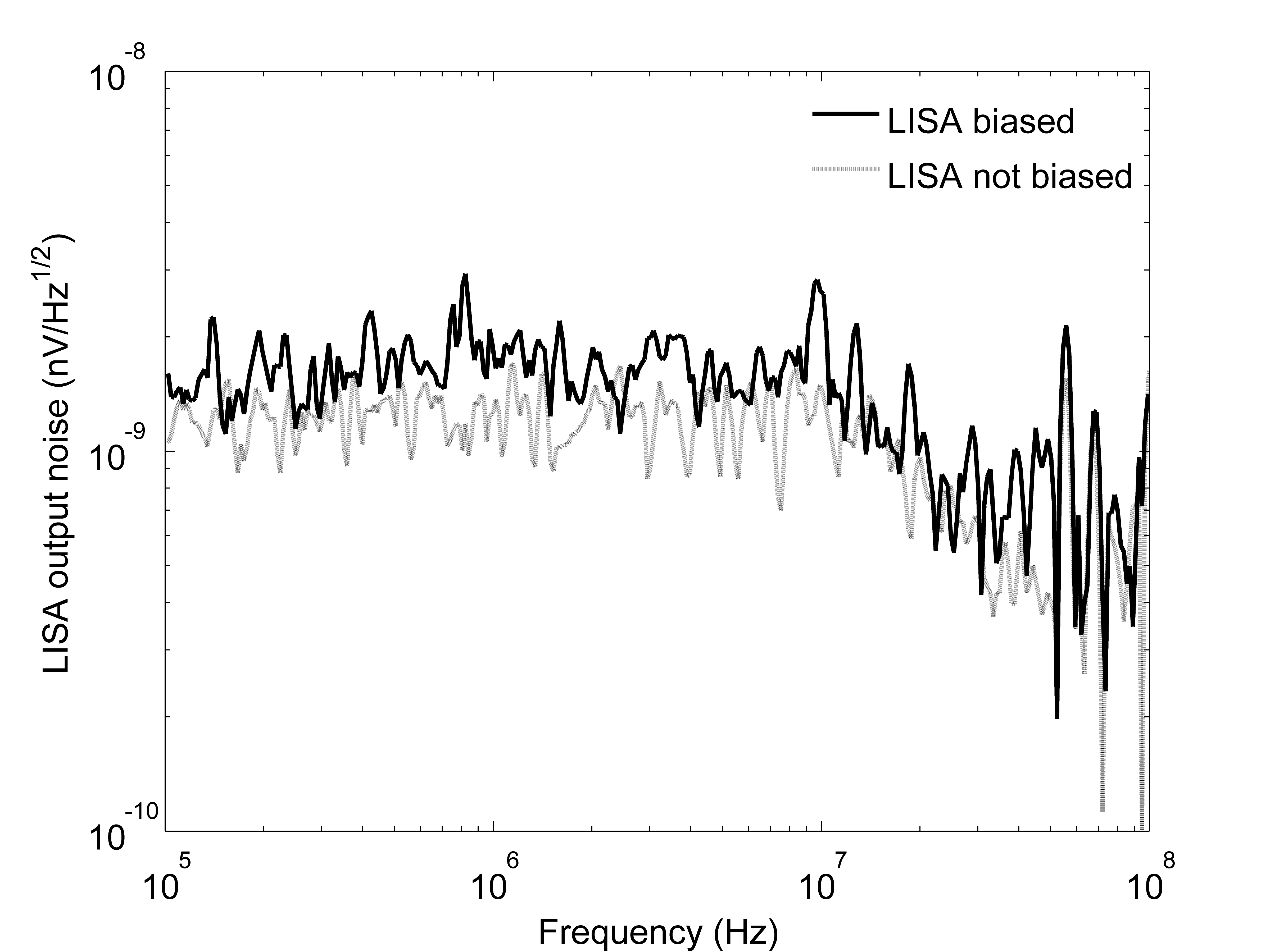}}
 \caption{\label{FIG:LISAnoise} Measured output voltage noise of the LISA with and without junction bias current.}
\end{figure}

We coupled the output voltage of the LISA to a spectrum analyser and
measured the noise from 1 to 100 MHz with the junction bias current
on and off. The noise measured with the bias current off, 1.3
nV/\rtHz, measures the baseline noise of the room temperature
readout electronics. With the bias current on, the noise increases
by 25$\%$ to 1.65 nV/\rtHz. Figure \ref{FIG:LISAnoise} shows the
measured noise spectra with the LISA bias current on and off. The
expected combination of the Nyquist current noise from the feedback
resistor and the intrinsic current noise from the individual SQUID
devices predicts a total input voltage noise of 1.55 nV/\rtHz, which
is in agreement with the measured noise level. Referred to the input of the LISA, this measured noise is 8.25 pA/\rtHz. This noise is currently dominated by the room temperature electronics and by increasing the transimpedance of the LISA, the referred current noise can be decreased.

%--------------------------------------------------------------
\section{Conclusions}
\label{SEC:conclusions}
\paragraph{}
The performance of the prototype LISA is such that it meets the
requirements for reading out multiplexed TES bolometers. The
bandwidth is increased by an order of magnitude from current
SQUID electronics, allowing the multiplexed channel count to
increase by this same factor and the noise and dynamic range meet the design requirements. 
Finally, although the prototype consisted of three separate SQUID array chips, the underlying circuit is very simple. Future versions of this device can easily be integrated and manufactured on a single chip. The constituent SQUIDs should be designed to minimize the ratio $R_o/Z_o$ to maximize the closed loop gain of the LISA.

\section{Acknowledgements}
We thank J. Clarke, D. Doering, and P. Richards for useful discussions. We also thank LBNL engineers John Joseph and Chinh Vu for their work on the room temperature electronics. Work at LBNL is supported by the Director, Office of Science, Office of High Energy and Nuclear Physics, of the U.S. Department of Energy under Contract No. DE-AC02-05CH11231.

%--------------------------------------------------------------
%%%%% References %%%%%
%\bibliographystyle{plain}
\bibliography{MyCMB-References}   % bibliography data in MyCMB-References.bib

\end{document}